\documentstyle[prl,aps]{revtex}
\def\caja{\mathsurround=0pt}
\def\eqalign#1{\,\vcenter{\openup1\jot \caja
        \ialign{\strut \hfil$\displaystyle{##}$&$
        \displaystyle{{}##}$\hfil\crcr#1\crcr}}\,}
\begin{document}
\draft
\twocolumn[\hsize\textwidth\columnwidth\hsize\csname @twocolumnfalse\endcsname
\title {Spectral weight function
for the half-filled Hubbard model:
a singular value decomposition approach}
\author{
C.E. Creffield,
E.G. Klepfish, E.R. Pike and Sarben Sarkar}
\address{Department of Physics, King's College London, Strand,
London WC2R 2LS, UK}
\maketitle
\begin{abstract}
The singular value decomposition technique is used to reconstruct
the electronic spectral weight function for a half-filled Hubbard
model with on-site repulsion $U=4t$ from Quantum Monte Carlo data.
A two-band structure for the
single-particle excitation spectrum is found to persist
as the lattice size exceeds the spin-spin correlation length.
The observed bands are flat in the vicinity of the $(0,\pi),(\pi,0)$
points in the Brillouin zone,
in accordance with experimental data for high-temperature
superconducting compounds.

\end{abstract}
\vskip2pc]
\narrowtext
The study of the normal properties of the high-temperature
superconducting compounds and
the attempts to construct a microscopic theory of superconductivity
in these materials has been in large part dedicated to the investigation
of the Fermi surface structure and its dependence on the strong
electronic correlations and antiferromagnetic order. Angle-resolved
photoemission experiments (ARPES) have recently provided a comprehensive
mapping of the Fermi surface in some of the cuprates (YBa$_2$Cu$_4$O$_2$ and
Bi$_2$Sr$_2$CaCu$_2$O$_{8+x}$)[\ref{Gofron}],[\ref{Dessau}],[\ref{Aebi}].
These results are particularly relevant to the high-temperature
superconductivity scenario in which
the phenomenon is related to the presence of
van Hove singularities in the density of states close
to the Fermi surface [\ref{Gofron}].
Quantum Monte Carlo simulations have recently
been used to compare these experimental
results with
numerical calculations in models
of strongly correlated electrons [\ref{DagottoNazarenko}],[\ref{Shadow}].
An essential part of these numerical results was the
extraction of the spectral weight function (SWF) for
single-particle excitations from the Quantum Monte Carlo
data calculated for imaginary time.

The SWF
is obtained as a solution of the following inverse
problem [\ref{ScalapinoSchutter}]:
\begin{equation}\label{inverse}
G(\vec k,\tau)=\int_{-\infty}^\infty{d\omega
{{e^{-\omega\tau}}\over {1+e^{-\omega\beta}}}A(\vec k,\omega)};
\hskip 8pt\,\, (0<\tau\leq\beta )
\end{equation}
where $G(\vec k,\tau)$ is the Matsubara Green's function for lattice
momentum $\vec k$ and
imaginary-time separation $\tau$, obtained in the
finite-temperature Monte Carlo simulations at temperature $1/\beta$, and
$A(\vec k,\omega)$ is the corresponding
SWF.

This inverse problem
admits an infinite class of solutions for $A(\vec k,\omega)$ which
will satisfy Eq.(\ref{inverse}) within small perturbations of the
l.h.s., originating from the statistical noise on the simulation
data. In recent works the
one-electron SWF for the single-band 2D Hubbard model was obtained from
Monte Carlo data using a maximum entropy approach
[\ref{WhiteVekic}],[\ref{Scalapinohalf-filling}],[\ref{Scalapinodoped}].
The finite-size effects evident in the data itself,
as well as in the reconstruction of the SWF,
led the authors of Ref.[\ref{WhiteVekic}] to conclude that
the pseudogap in the single-particle spectrum
observed in small clusters (up to $8^2$) is a lattice-size
artifact which disappears as the size of the simulated system increases.
They suggested therefore that for on-site repulsion
weaker than the Hubbard bandwidth ($8t$)
the antiferromagnetic
correlation length would be comparable to
the size of a sufficiently small
lattice and would effectively create long-range antiferromagnetic
order.
Based on this conclusion the $U=4t$ coupling
(where $U$ is the on-site Coulomb repulsion
and $t$ is the nearest-neighbour hopping parameter)
was regarded as belonging to
the weak-coupling regime, where the band structure is essentially similar to
that of noninteracting electrons for all temperatures above
zero, in accordance with the Mermin-Wagner theorem. Following this argument,
a gap originating from spin-density waves
(SDW) will be both temperature and finite-size sensitive.
The low-temperature simulations, as presented in [\ref{White92}]
for $\beta=12$ show indeed a stable gap even for relatively large
lattice size. Such temperature dependence of the gap would be expected
to correspond to a strong temperature dependence of the spin-spin correlations.
However, as the numerical results of Refs.[\ref{Moreo93}] and
[\ref{FanoOrtolaniParola}] show, the spin-spin correlations
at $U=4t$ and $4^2$ lattice fall
rapidly within 2 lattice spacings with only minor dependence
on the temperature in the range $\beta=4-12$. The SWF
presented in Ref.[\ref{WhiteVekic}] shows a clear two-peak structure
for this value of $U$ and lattice linear dimension larger than the correlation
length.
Hence the possibility that the band structure is
due to the short-range antiferromagnetic
order (as was suggested in Ref.[\ref{Aebi}])
demands an accurate mapping of the
on-site repulsion regimes for coupling $U<8t$.

A detailed study of the SWF as a solution of
an inverse problem requires a quantitive criterion for the
resolution limits of the reconstruction technique.
Without such a criterion an existing gap might
be overlooked in the reconstruction procedure.
Moreover, the single-particle spectrum derived
from the distribution of the function $A(\vec k,\omega)$ along the Brillouin
zone is based on the identification of the peaks of this
function. Thus the resolution of the two-peak structure,
even if there is no clear evidence of the vanishing of this function
between the peaks,
is important.

In our work we use a method based on the singular value
decomposition (SVD) approach widely used in the field of
inverse problems [\ref{Pike}]. This technique allows us to
define
the resolution limits quantitatively, thus providing an upper bound on the size
of the gap with respect to the lattice size.

The work presented is an investigation performed
on lattice sizes ranging between $4^2$ and $12^2$.
The simulation temperature was chosen as $\beta=5$
and on-site repulsion as $U=4t$.
To avoid the sign problem the model was simulated at half-filling. However,
the SVD treatment can be extended to finite doping, as well as to
other Green's functions.

Eq.(\ref{inverse}) can be rewritten in operator form
as
follows [\ref{Pike}]:
\begin{equation}\label{inverse_new}
G(\tau_i)=({\cal K} A)(\tau_i).
\end{equation}
${\cal K}$ is
the integral operator of the r.h.s. of (\ref{inverse}), with
the variable $\tau$ in the kernel discretised, acting on the
function $A$. This operator defines a transformation from the $L^2$
space of the SWF
to the data space of which $G(\tau_i)$ is a vector.
This transformation may be regarded as a rectangular
matrix, ${\cal K}$, one of whose dimensions is infinite and the other
equal to the dimension of the data space, namely to
the number of data points $n_\tau$. A conjugate operator ${\cal K^\ast}$
acts in the data space and is determined by the following equality of inner
products:
\begin{equation}\label{conjugate}
\sum_{i,j=1}^{n_\tau}W_{\tau_i\tau_j}
({\cal K} A)(\tau_i)G(\tau_j)
=\int d\omega A(\omega)({\cal K^\ast}G)(\omega),
\end{equation}
with a metric tensor $W$ which can account for
the correlations in the data points. A common choice is to
define this metric
as the inverse of the covariance matrix
[\ref{Pike}].
The operator ${\cal K K^\ast}$
is represented by a symmetric square matrix
with a positive set of $n_\tau$ eigenvalues $\{\alpha_i^2\}$ and
corresponding set of orthonormal eigenvectors $\{v_i\}$.
These vectors form a basis in the data space.
The operator ${\cal K^\ast K}$ acting in the space to which
the solution $A(\omega)$ belongs has a set of $n_\tau$ orthonormal
eigenfunctions $\{u_i\}$ with the same eigenvalues.
If the set $\{\alpha_i\}_{i=1}^{n_\tau}$ is arranged in descending
order each vector $v_k$ and function $u_k$ will have $k-1$ nodes.
These two sets $\{u_i\}$ and $\{v_i\}$ and the set
$\{\alpha_i\}_{i=1}^{n_\tau}$
form a singular system satisfying the equations:
\begin{equation}\label{svdset}
\eqalign{
{\cal K }u_k&=\alpha_k v_k \cr
{\cal K^\ast}v_k&=\alpha_k u_k.}
\end{equation}
The matrix representation of the operator ${\cal K K^\ast}$ is
obtained[\ref{Pike}] from the kernel of Eq.(\ref{inverse}) as:
\begin{equation}\label{squarmatrix}
({\cal K K^\ast})_{\tau_m\tau_n}=\sum_{\tau_k}
{W_{\tau_m\tau_k}\int_{-\infty}^\infty d\omega
{{e^{-\omega(\tau_k+\tau_n)}}
\over {(1+e^{\omega\beta})^2}}}.
\end{equation}

The solution of Eq.(\ref{inverse}) expanded in the functions $\{u_k\}$
is given by:
\begin{equation}\label{solut}
A(\omega)=\sum_{k=1}^{n_\tau}{1\over {\alpha_k}}<G,v_k>u_k(\omega)
\end{equation}
where $<,>$ denotes the inner product in the data space.
The smallest singular values
will
greatly amplify the contribution of the statistical noise to the
expansion coefficients in Eq.(\ref{solut}).
We therefore
introduce a cut-off into the expansion to exclude terms
with ${\alpha_k\over {\alpha_1}}$ smaller than the statistical error in
$G(\tau)$ [\ref{Pikeregularisation}].

The ``resolution ratio" depends on the number of nodes in the function
with the highest index $k_{max}$
included
in the truncated expansion [\ref{Pikeresolution}].
Here prior information about the
support of the solution is essential.
Since we assume that the $A(\omega)$ is localised within
a certain range of $\omega$, we confine the singular functions to this
range.
This is done by multiplying
the kernel in Eq.(\ref{inverse})  by a profile function which is approximately
1 within the support and smoothly vanishes outside it [\ref{Pikeresolution}].
The rate of the decrease of the singular values for
the limited support grows as the profile function gets tighter, therefore,
the truncation in (\ref{solut}) will have a smaller $k_{max}$. However,
since the support of the singular functions is limited, their
nodes will be more closely spaced, thus giving better resolution.
To investigate the resolution limits, artificial
data
was generated by substituting an $A(\omega)$ consisting of two
Gaussian peaks into Eq.(\ref{inverse}).
The inverse problem was then
solved for $A(\omega)$ with the expansion (\ref{solut})
truncated at ${{\alpha_1}\over {\alpha_k}}$ exceeding
a hypothetical noise level. Similar to the results of
Ref.[\ref{Pikeresolution}]
we
found that for a fixed noise the resolution
improves as the range of support decreases.

The reconstruction of the SWF for $\vec k$ at the vicinity
of $\vec k=({\pi\over 2}, {\pi \over 2})$
is given in Fig.1. It can be seen that there are negative
side-lobes.
Just as a positive $\delta-$function, imaged over a limited Fourier
bandwidth, turns into Airy rings which have negative side-lobes,
we must expect a localised SWF, reconstructed over a limited singular
function bandwidth, to have similar features.
A regularization procedure which would restrict the solution to
positive values only may also reduce the resolution of
its reconstruction [\ref{Pikeregularisation}].
We present here the results of the $8^2$-$12^2$ lattices, since the
existence of the gap in the smaller lattices is beyond controversy.
Limiting the support of $A(\vec k, \omega)$ to the range
$[-8,8]$, allowed
us to identify a gap unseen in an infinite-support reconstruction.
According to our statistical noise we truncated the expansion of
$A(\vec k,\omega)$
with the ratio ${{\alpha_1}\over {\alpha_k}}$
not exceeding $100$ (for a 1\% statistical error as estimated
for 4000 Monte Carlo measurements)
which allowed us to
use 9 singular functions in the unlimited-support case and 7 in
the reconstruction with the finite profile.

Since the statistical noise determines the level of truncation of the
singular-function expansion it is clear that at a particular
truncation level, not all the data included in the $n_\tau$ points
are of identical relevance.  In accordance with the
application of sampling theory in other inverse problems
we identify as essential the nodes
of the $v_{k_{max}}$.
This is a generalisation of the Nyquist rate in Fourier theory.
For example, in the inversion of the Laplace transform
[\ref{Sampling}] the optimal sampling was found to be exponential
which corresponds to the nodes of $v_{k_{max}}$. Application of this
sampling technique to the inverse problem (\ref{inverse})
is performed as follows: we estimate
$k_{max}$ from the knowledge of the noise and the spectrum of the singular
values. Then the data is evaluated at points nearest to the nodes
of $v_{k_{max}}$ and the SVD reconstruction is done for these points only.
This procedure leads to an obvious advantage from the computational
point of view, since the evaluation of the Matsubara Green's
functions is reduced to $1\over 5$ of the total points. Moreover,
by
separating the data points by a large spacing we reduce the correlations within
the data and the metric on the data space can be taken as Euclidean,
which also substantially reduces the calculational effort.
For the $8^2$ lattice
we evaluated the SWF using 7 data points only and compared the results
with that obtained from using 40 equally spaced points.
For the latter case we accounted
for the correlations using the covariance matrix as the metric
in the data space. The results show remarkable agreement, thus fully justifying
the sampling approach in this problem.

In Fig.2 we present the density of states calculated for the three biggest
lattice sizes as the
solution $N(\omega)$ of:
\begin{equation}\label{inverseg}
\tilde G(0,\tau)=\int_{-\infty}^\infty{d\omega
{{e^{-\omega\tau}}\over {1+e^{-\omega\beta}}}N(\omega)}
\end{equation}
with $\tilde G(0,\tau)$ being the zero-space-separation Matsubara
Green's function. In this reconstruction the range of the support
for $N(\omega)$ is [-12,12].

The gap decreases as the lattice size increases,
in correspondence with the results of Veki\'c and White [\ref{WhiteVekic}],
but it is still clear even in the $12^2$ lattice. In fact the
difference in the size and depth of the gap is very small as
the lattice size increases from $8^2$ to $12^2$, while the
linear dimension of the lattice is considerably larger than the SDW
correlation length.
The existence of two clear peaks in the density of states (as well
as in $A(\vec k_{Fermi},\omega)$) suggests a two-band structure of
the spectrum.

The reliability of our results has been tested against the following
criteria: the successful reconstruction of the Monte-Carlo data
when the solution of $A(\vec k, \omega)$ is substituted in the r.h.s.
of Eq.(\ref{inverse}); the successful reconstruction of the same data by:
\begin{equation}\label{reconseigen}
G=\sum_k^{k_{max}}<G,v_k>v_k;
\end{equation}
and, finally, by checking the first three moments of the SWF:
\begin{equation}\label{moments}
\mu_n=\int\omega^n A(\vec k,\omega) d\omega \,\,\,\,(n=0,1,2)
\end{equation}
and the same moments for the density of states.
We present in Table 1 the values of these moments for the density of states,
showing excellent agreement with the analytic values 1,0,8 for
the zeroth, first and second moments respectively [\ref{WhiteVekic}].
\begin{center}
\begin{tabular}{|c|c|c|c|c|c|c|c|}
\hline
    \multicolumn{1}{|c|}{lattice} & \multicolumn{1}{c|}{$\mu_0$}
& \multicolumn{1}{c|}{$\mu_1$}
&\multicolumn{1}{c|}{$\mu_2$} \\
\hline
 $8^2$ & 1.00
&0.00
& 7.99 \\
\hline
 $10^2$ &1.00
&0.00
& 7.79\\
\hline
 $12^2$ & 1.00
& 0.00
& 7.77 \\
\hline
\end{tabular}
\smallskip
\newline\noindent
Table 1. The first three moments of the density of states versus lattice size.
\end{center}
Fig.3 shows
the single-particle spectrum derived from identifying $E(\vec k)$ as the
location of a peak of $A(\vec k,\omega)$, where the values of the
momentum $\vec k$ scan over the Brillouin zone.
In the calculation of the spectrum we observe a significant
difference between the $8^2$
and the $12^2$ results. Both lattices exhibit a clear two-band structure of
the single-particle spectrum. However, while the former can be fitted well
with the mean-field SDW formula:
\begin{equation}\label{SDWspectrum}
E(\vec k)_{MF}=\sqrt{(\epsilon(\vec k)^2+\Delta^2)}
\end{equation}
where $\epsilon(\vec k)=-2t(\cos{k_x}+\cos{k_y})$ and $\Delta=0.64t$,
the $12^2$ lattice shows a bigger gap at the corners of the
noninteracting Fermi surface than at the point $({\pi\over 2}, {\pi\over 2})$.
Indeed the $12^2$ lattice simulation gives a smaller gap at this point than
the $8^2$ and the $10^2$ (see Fig.1). However, at the the points $(\pi,0)$
the gap remains persistently clear for all the lattices.
This result can be interpreted as suggesting a gap due to an anisotropic
pairing channel [\ref{Dessau}] (although
not necessarily of $d_{x^2-y^2}$ symmetry). Higher resolution
calculations, based on data with better statistics
will be required to examine
this possibility.
We also observe the feature of a flat band near $\vec k = (\pi, 0)$ similar
to the results of Dagotto {\it et al.}[\ref{DagottoNazarenko}] and the
experimental data from ARPES [\ref{Dessau}].
This flatness may lead to a van Hove-like singularity in the density of states.

The SVD method brings to this inversion problem a more rigorous and
controlled approach than the maximum entropy technique.
Contrary to previous findings
the results of our calculations suggest that $U=4t$ is in a regime in which
the spectrum of quasiparticle (damped) excitations has a two-band structure,
in spite of the on-site repulsion being sufficiently weak to
allow substantial double occupancy.
There is some support for the relevance of $U=4t$ regime to
high T$_c$ superconductivity from a recent work
by Beenen and Edwards [\ref{BeenenEdwards}].

We thank J. Jefferson, A. Bratkovsky, D. Edwards and P. Kornilovitch
for stimulating discussions. This work was suported by SERC grant
GR/J18675 and our general development of SVD techniques by
the US Army Research Office, agreement no. DAAL03-92-G-0142.
\vfill
\newpage
\noindent{\bf References}
\begin{enumerate}
\item\label{Gofron}{K. Gofron {\it et al., Phys. Rev. Lett.} {\bf 73}, 3302
(1994)}
\item\label{Dessau}{D.S. Dessau {\it et al., Phys. Rev. Lett.} {\bf 71}, 2781
(1993);\\
D.M. King {\it et al., Phys. Rev. Lett.} {\bf 73}, 3298 (1994).}
\item\label{Aebi}{P. Aebi {\it et al., Phys. Rev. Lett.} {\bf 72}, 2757
(1994)}
\item\label{DagottoNazarenko}{E. Dagotto, A. Nazarenko and M. Boninsegni,
{\it Phys. Rev. Lett.} {\bf 73}, 728 (1994).}
\item\label{Shadow}{S. Haas, A. Moreo and E. Dagotto, {\it unpublished}.}
\item\label{quasiparticle}{A. Moreo, S. Haas, A. Sandvik and E. Dagotto,
{\it unpublished}.}
\item\label{ScalapinoSchutter}
{H.-B. Sch\"uttler and D.J. Scalapino, {\it Phys. Rev.} {\bf B34}, 4744
(1986).}
\item\label{WhiteVekic}
{S.R. White, {\it Phys. Rev.} {\bf B44}, 4670 (1991);\\
M. Veki\'c and S.R. White, {\it Phys. Rev.} {\bf B47}, 1160 (1993).}
\item\label{Scalapinohalf-filling}
{N. Bulut, D.J. Scalapino and S.R. White, {\it Phys. Rev. Lett.} {\bf 73},
748 (1994).}
\item\label{White92}{S.R. White, {\it Phys. Rev.} {\bf B46}, 5678 (1992).}
\item\label{Moreo93}{E. Dagotto, {\it Rev. Mod. Phys.} {\bf 66}, 763 (1994).}
\item\label{FanoOrtolaniParola} {G. Fano, F. Ortolani and A. Parola,
{\it Phys. Rev.} {\bf B46}, 1048 (1992)}
\item\label{Scalapinodoped}
{N. Bulut, D.J. Scalapino and S.R. White, {\it Phys. Rev. Lett.} {\bf 72},
705 (1994).}
\item\label{Pike}{M. Bertero, C. De Mol and E.R. Pike, {\it Inverse Problems}
{\bf 1}, 301 (1985);\\
M. Bertero and E.R. Pike, {\it Handbook of Statistics} {\bf 10},
{\it Eds. N.K. Bose and C.R. Rao}, Elsevier Science
Publishers (1993).}
\item\label{Pikeregularisation}
{M. Bertero, P. Branzi, E.R. Pike and  L. Rebolia, {\it Proc. R. Soc. London}
{\bf A 415}, 257 (1988);\\
M. Bertero, C. De Mol and E.R. Pike, {\it Inverse Problems}
{\bf 4}, 573 (1988).}
\item\label{Pikeresolution}
{M. Bertero, P. Boccacci and E.R. Pike, {\it Proc. R. Soc. London}
{\bf A 383}, 15 (1982);
{\bf A 393}, 51 (1984);\\
M. Bertero, P. Branzi and E.R. Pike, {\it Proc. R. Soc. London}
{\bf A 398}, 23 (1985).}
\item\label{Sampling}
{M. Bertero and E.R. Pike, {\it Inverse Problems} {\bf 7}, 1 (1991);
21 (1991).}
\item\label{BeenenEdwards}{J. Beenen and D.M. Edwards,
to be published in {\it Jour. of Low Temp. Phys. .}}
\end{enumerate}
\noindent{\bf Figure captions}
\newline\noindent Fig.1a: $A(\pi/2,\pi/2,\omega)$ for a $8^2$ lattice.
Solid line -- SVD solution with infinite support, dotted line -- SWF
support is limited in the range [-8,8].
\newline\noindent Fig.1b: $A(2\pi/5,3\pi/5,\omega)$ for a $10^2$ lattice.
Solid line -- SVD solution with infinite support, dotted line -- SWF
support is limited in the range [-8,8].
\newline\noindent Fig.1c: $A(\pi/2,\pi/2,\omega)$ for a $12^2$ lattice.
Solid line -- SVD solution with infinite support, dotted line -- SWF
support is limited in the range [-7,7].
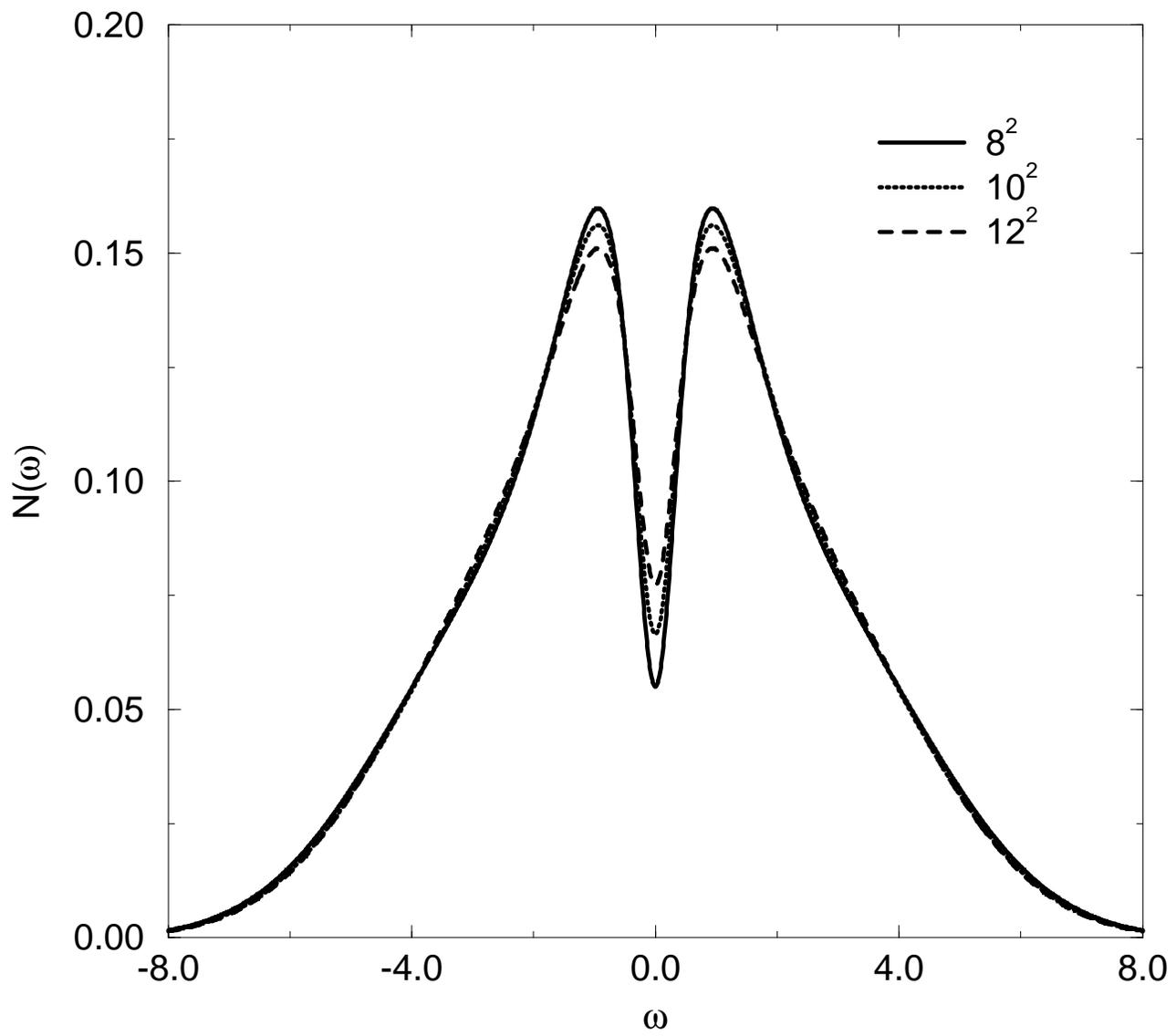
\newline\noindent Fig.2: Density of states as a function of $\omega$.
\newline\noindent Fig.3a: $8^2$ lattice. Spectrum of single-particle
excitations. Lower band -- stars; upper band -- circles. Dotted line
-- $\epsilon(\vec k)= -2t(\cos{k_x}+\cos{k_y})$; solid line --
$E(\vec k)_{MF}=\pm \sqrt{\epsilon(\vec k)^2+\Delta^2}$.
Points in the Brillouin zone:
$\Gamma $ -- $(0,0)$; $\bar{\rm M}$ -- $({\pi\over 2},{\pi\over 2})$;
M -- $(\pi,\pi)$; X -- $(\pi,0)$.
\newline\noindent Fig.3b: Same as (a) for a $12^2$ lattice.
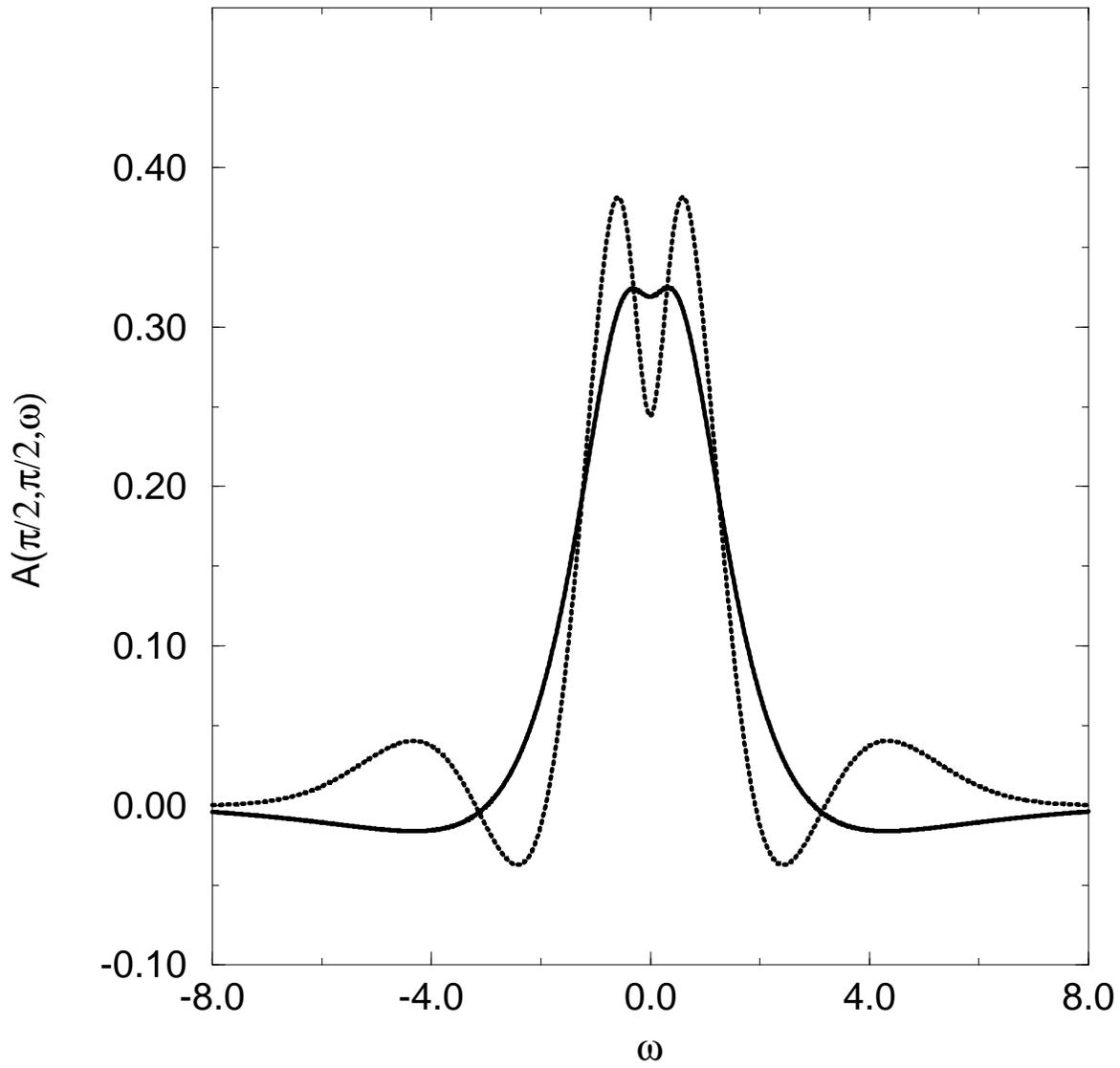
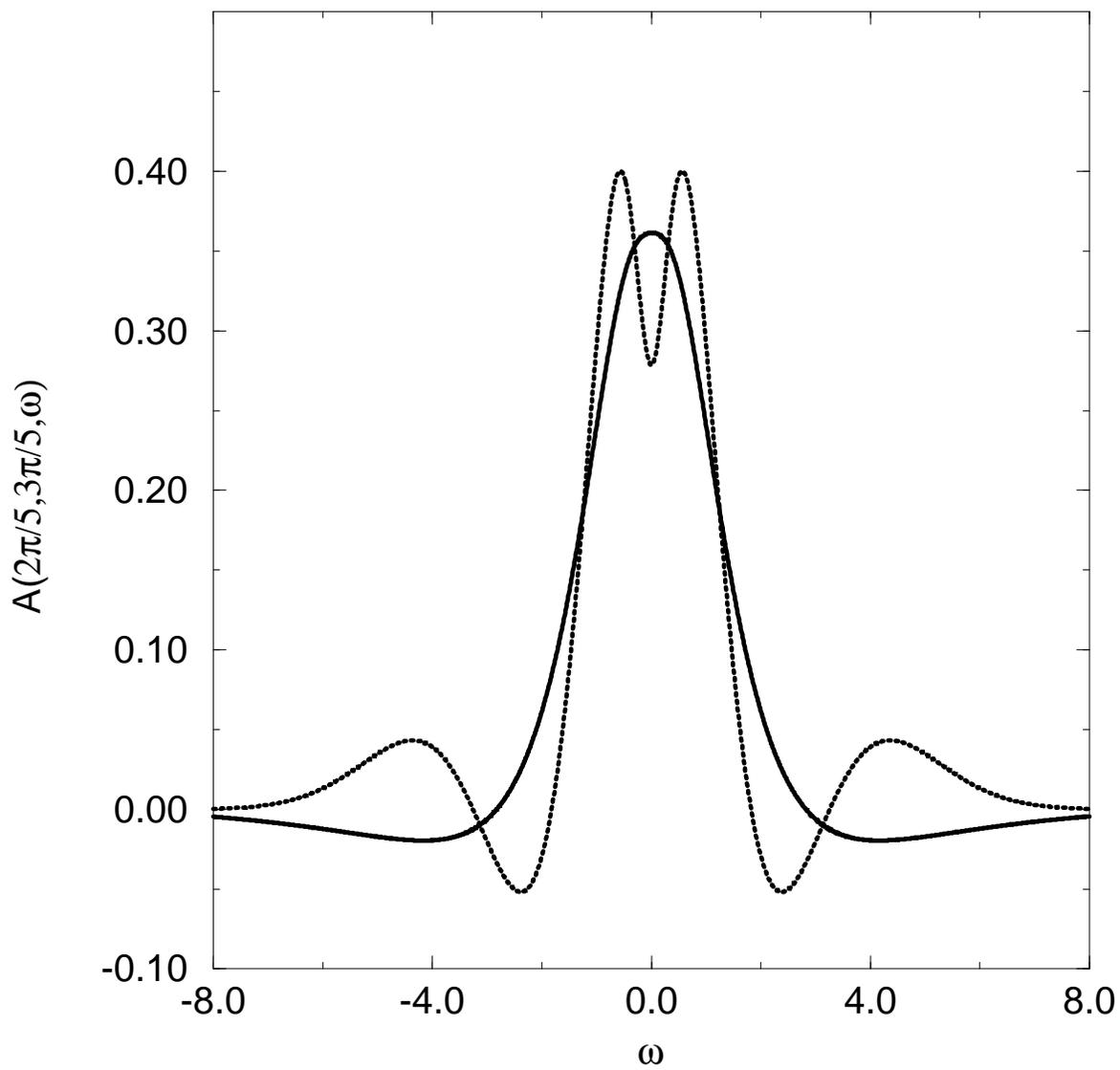
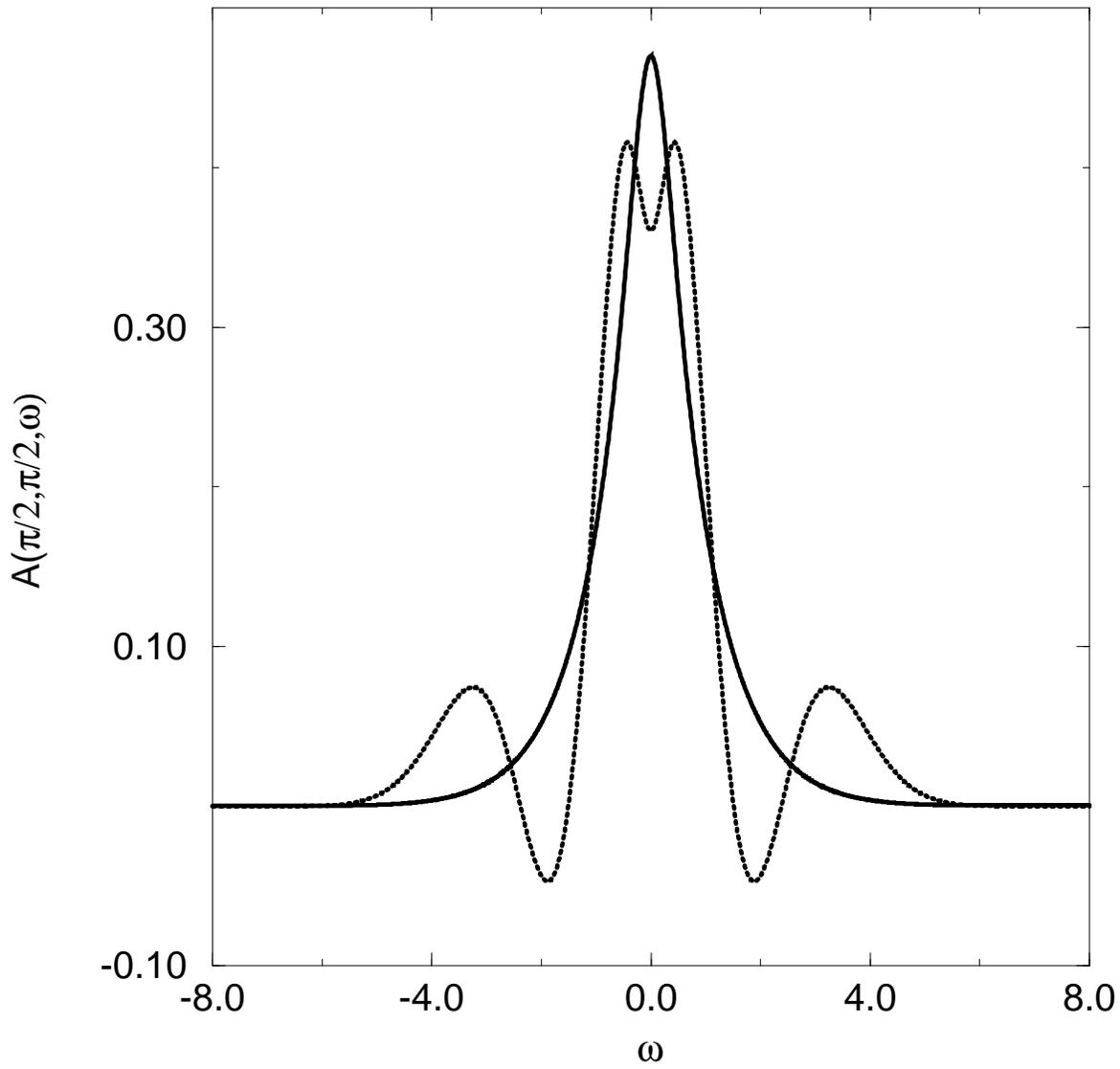
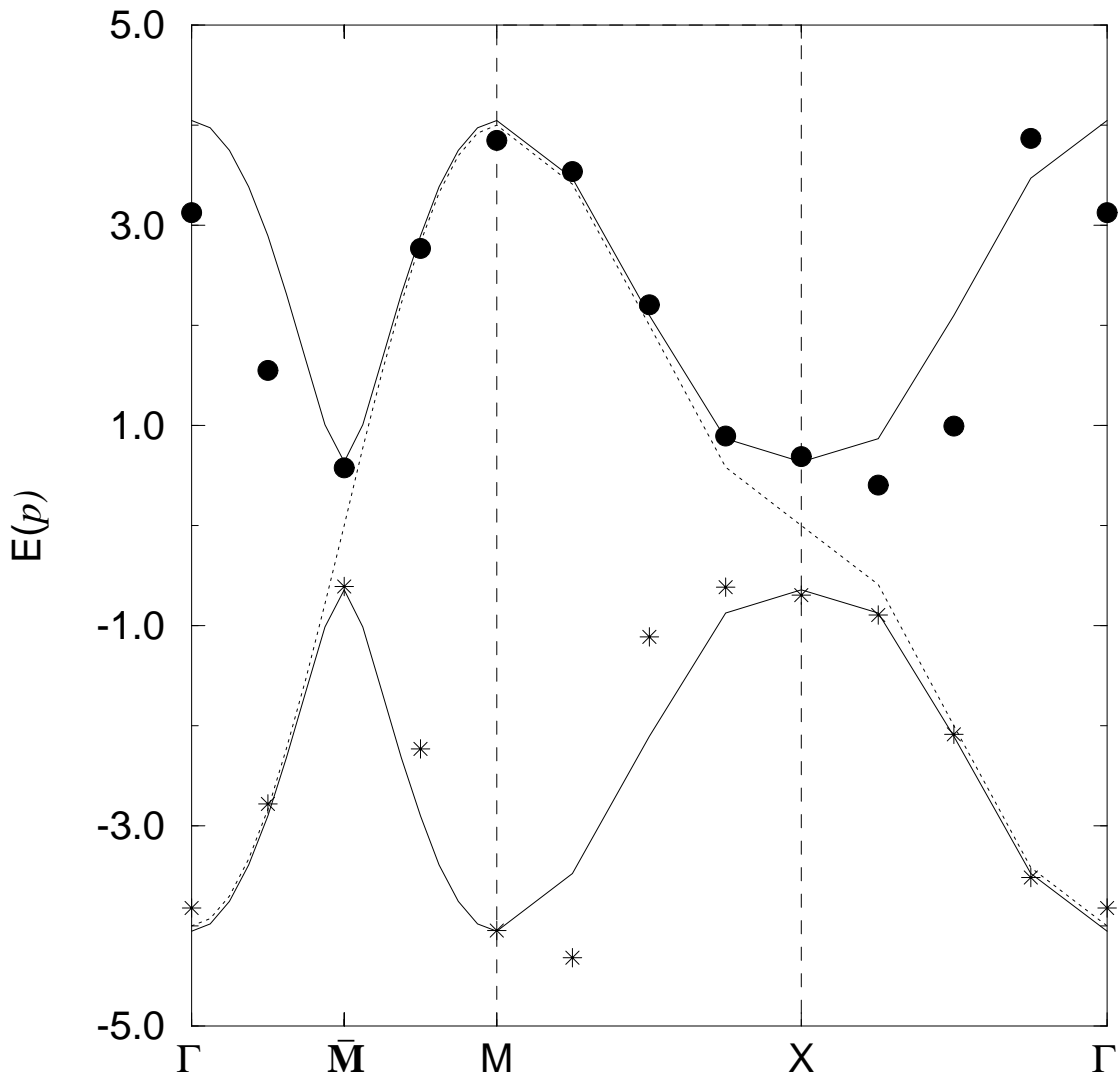
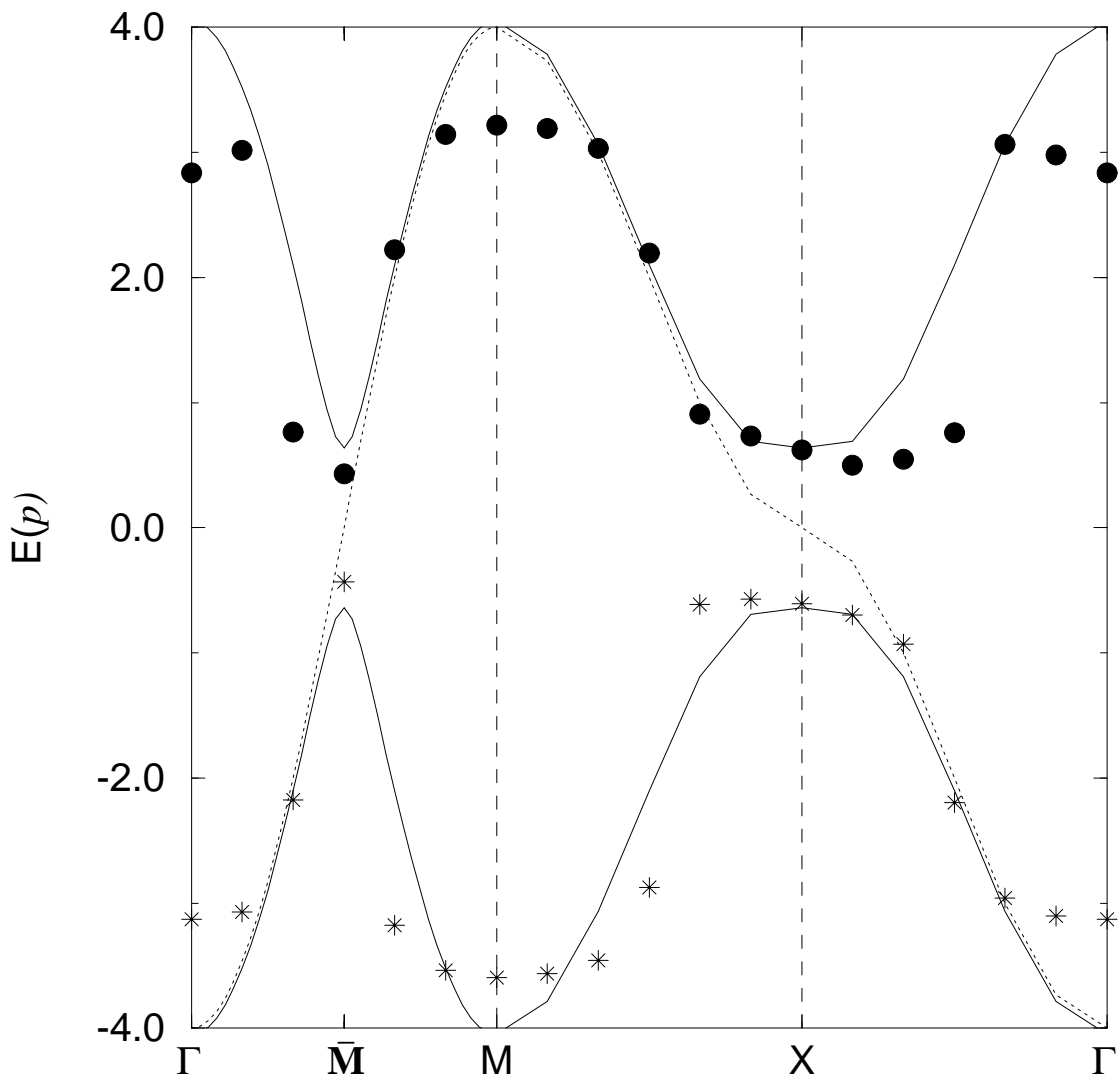
\end{document}